\documentclass[graphics, twocolumn, usenatbib]{mn2e}
\usepackage[utf8]{inputenc}
\usepackage{times}
\usepackage{natbib} 
\usepackage{epsfig} 
\usepackage{graphicx} 
\usepackage{color}
\usepackage{aas_macros} 
\usepackage{amssymb}
\usepackage{amsmath}
\usepackage[title]{appendix}
\usepackage{hyperref}	
\hypersetup{colorlinks=true,linkcolor=blue,citecolor=blue,filecolor=blue,urlcolor=blue}
\usepackage[caption=false]{subfig}

\newcommand{\enzo}{\texttt{Enzo~}}

\newcommand{\mpch} {\rm $h^{-1}$ Mpc\,\,} 
 
\newcommand{\msolar} {$\rm{M_{\odot}}~$}
\newcommand{\msolarc} {$\rm{M_{\odot}}$}

\newcommand{\zsolarc} {$\rm{Z_{\odot}}$}

\newcommand{\molH} {$\rm{H_2}$~}
\newcommand{\molHc} {$\rm{H_2}$}

\newcommand{\JLW} {J$_{LW}$}

\newcommand{\rarepeak} {\textit{Rarepeak~}}
\newcommand{\normal} {\textit{Normal~}}
\newcommand{\void} {\textit{Void~}}
\newcommand{\voidc} {\textit{Void}}

\begin{document}
\title{The Emergence of the First Star-free Atomic Cooling Haloes in the Universe}
\author[J. A. Regan, J. H. Wise,  B.W. O'Shea \&  M.L. Norman]{John A. Regan$^{1}$\thanks{E-mail:john.regan@dcu.ie}, John H. Wise$^{2}$ Brian W. O'Shea$^{3,4,5,6}$ \& Michael L. Norman$^7$\\ $^1$Centre for Astrophysics \& Relativity, School of Mathematical Sciences,
         Dublin City University, Glasnevin, D09 W6Y4, Ireland\\$^2$Center for Relativistic Astrophysics, Georgia Institute of Technology, 837 State Street, Atlanta, GA 30332, USA\\
         $^3$National Superconducting Cyclotron Laboratory, Michigan State
         University, MI, 48823, USA\\
         $^4$Department of Physics and Astronomy, Michigan State University,MI, 48823, USA\\
         $^5$Department of Computational Mathematics, Science and Engineering, Michigan State University, MI, 48823, USA\\    
         $^6$Joint Institute for Nuclear Astrophysics - Center for the Evolution of the Elements, USA\\
         $^7$Center for Astrophysics and Space Sciences, University of California, San Diego, 9500 Gilman Dr, La Jolla, CA 92093\\}

\pubyear{2018}
\label{firstpage}
\pagerange{\pageref{firstpage}--\pageref{lastpage}}
\maketitle

\begin{abstract}
Using the Renaissance suite of simulations we examine the emergence of pristine atomic cooling haloes that are both metal-free and star-free in the early Universe. The absence of metals prevents catastrophic cooling, suppresses fragmentation, and may allow for the formation of massive black hole seeds. Here we report on the abundance of pristine atomic cooling haloes found and on the specific physical conditions that allow for the formation of these direct-collapse-black-hole (DCBH) haloes. In total in our simulations we find that 79 DCBH haloes form before a redshift of 11.6. We find that the formation of pristine atomic haloes is driven by the rapid assembly of the atomic cooling haloes with mergers, both minor and/or major, prior to reaching the atomic cooling limit a requirement. However, the ability of assembling haloes to remain free of (external) metal enrichment is equally important and underlines the necessity of following the transport of metals in such simulations.
The candidate DCBH hosting haloes we find, have been exposed to mean Lyman-Werner radiation fields of J$_{LW}$ $\sim 1$ J$_{21}$ and typically lie at least 10 kpc (physical) from the nearest massive galaxy. Growth rates of the haloes reach values of greater than $10^7$ \msolar per unit redshift, leading to
significant dynamical heating and the suppression of efficient cooling until the halo crosses the atomic cooling threshold. Finally, we also find five synchronised halo candidates where pairs of pristine atomic 
cooling haloes emerge that are both spatially and temporally synchronised. 
\end{abstract}



\section{Introduction}\label{Sec:Introduction} 
Supermassive black holes (SMBHs) with masses upwards of a billion solar masses have been observed less than one billion years after the Big Bang \citep{Fan_2006, Mortlock_2011, Wu_2015, Banados_2018}. However, the mechanisms which allow for the formation of supermassive black holes is hotly debated and currently unknown \citep[for a recent review see][]{Woods_2018}. The 
mainstream scenarios fall into two main brackets. The first mechanism uses \textit{light} seeds as the 
origin for the massive black hole seeds. Light seeds are thought to have masses between 30 and 1000 \msolar masses and may be formed from the end point of Population III (PopIII) stars \citep{Abel_2002, Bromm_2002, Madau_2001}. Light seeds may also evolve from the core collapse of a dense stellar cluster \citep{Begelman_78b, Freitag_2006, Merritt_2008, Devecchi_2008, Freitag_2008, Lupi_2014, Katz_2015} where stellar collisions result in the formation of a massive black hole. However, there is a general consensus within the community that 
growing from light seed masses up to one billion solar masses may be demanding in the early Universe and that the vast majority of light seeds suffer from starvation in their host halo \citep{Whalen_2004, Alvarez_2009, Milosavljevic_2009, Smith_2018}; however, see \citep{Alexander_2014,Inayoshi_2018, Pacucci_2017} for examples of super-Eddington accretion mechanisms which may circumvent light seed growth restrictions. 

The second mechanism advocates for \textit{heavy} seeds with initial masses between 1000 \msolar and 100,000 \msolarc. This scenario is commonly referred to as the "Direct Collapse Black Holes" (DCBH) scenario \citep{Eisenstein_1995b, Oh_2002, Bromm_2003} and relies on the direct collapse of a metal-free gas cloud directly into a massive black hole. Depending on the 
exact thermodynamic conditions of the collapse the massive black hole phase may be preceded by an intermediary stage involving a super-massive star \citep{Shapiro_1979, Schleicher_2013, Hosokawa_2013, Inayoshi_2014, Woods_2017, Haemmerle_2017b, Haemmerle_2017} or a quasi-star \citep{Begelman_2006, Begelman_2008}. Initial numerical investigations of the collapse of atomic cooling haloes revealed that the collapse could proceed monolithically and that the formation of a massive black hole seed with a mass up to 100,000 \msolar masses was viable in the early Universe where atomic cooling haloes were both metal-free and free of \molH \citep{Bromm_2002, Wise_2008a, Regan_2009b, Regan_2009}. 

As the numerical investigations became more sophisticated, the research landscape shifted to understanding how metal-free atomic cooling haloes could exist which remained free of rampant star formation. \molH cooling within mini-haloes, which would precede atomic cooling haloes, would lead to the formation of PopIII stars thus shutting off the pathway to massive black hole seed formation. 
\molH can be dissociated by radiation in the Lyman-Werner (LW) band \citep{Field_1966} between 11.8 and and 13.6 eV. If the intensity of LW radiation is strong enough then \molH formation can be suppressed, allowing for the formation of an atomic cooling halo in which \molH cooling is prevented and the halo must cool and collapse on the so-called atomic track. A number of authors 
\citep{Shang_2010, Wolcott-Green_2011, Sugimura_2014, WolcottGreen_2012, Regan_2014a, Visbal_2014, Agarwal_2015a, Latif_2015}
examined the intensity of LW radiation required to completely suppress \molH formation and found that the intensity of LW radiation impinging onto a nascent halo needed to be upwards of 1000 J$_{21}$\footnote{J$_{21}$ is shorthand for $1 \times 10^{-21} \ \rm{erg \ s^{-1} cm^{-2} Hz^{-1} sr^{-1}}$ and measures the intensity of radiation at a given point.}. Only pristine and metal-free haloes in close proximity to another rapidly star-forming halo would be able to fulfill that criteria given that the 1000 J$_{21}$ value is orders of magnitude above expected mean background values \citep[e.g.][]{Ahn_2009}. Two haloes developing closely separated in both time and space would allow for this mechanism and hence the "synchronised-halo" model was developed by \cite{Dijkstra_2008}, which advocated this approach as being conducive to the formation of atomic cooling haloes that allow the full suppression of \molHc. \cite{Regan_2017} tested the theory rigorously through numerical simulations, showing that atomic cooling haloes that develop and are sub-haloes of one another can lead to the complete suppression of \molH in one of the haloes and hence to an isothermal collapse of the core of one of the pair. The exact abundance of synchronised haloes is challenging to predict analytically and even in optimistic evaluations the number density of synchronised pairs may only be able to seed a sub-population of all SMBHs \citep{Visbal_2014b, Inayoshi_2015b, Habouzit_2016}. 

More recently \cite{Wise_2019}, hereafter W19, showed that the rapid assembly of haloes can also lead to the suppression of \molH and should be significantly more common than the synchronised pair scenario (though this 
mechanism does not necessarily lead to pure isothermal collapse while the synchronised scenario should). Dynamical heating \citep{Yoshida_2003a, Fernandez_2014} can suppress the impact of \molH cooling, thus keeping an assembling halo hotter and preventing the formation of stars. W19 investigated two haloes in particular from a set of high resolution adaptive mesh refinement simulations of the early Universe that they found had breached the atomic cooling limit, were metal-free and had not formed stars. The two haloes that they targeted for detailed examination were the most massive halo (MMHalo) and the most irradiated halo (LWHalo) at the final 
output of the simulation, redshift 15. W19 found that the haloes were subject to only relatively mild LW exposure and that in the absence of all other external effects should have formed stars. They found that the haloes experienced especially rapid growth compared to typical haloes and that the extra dynamical heating effects driven by the rapid growth allowed the haloes to remain star-free. Their examinations also showed that the haloes did not show any initial signs of rapid collapse - however they did not run their simulations beyond the formation of the first density peak and further evolution of these haloes is still required to determine the detailed characteristics of the objects that form. In this study we examine the entire dataset of metal-free and star-free haloes produced by the simulations used in W19. As such, this study is more comprehensive and allows for a broader analysis of the physics driving the formation of these pristine objects. The goal of this study is to look at the Renaissance simulation dataset in its entirety. Here we identify DCBH candidates at each redshift and also investigate the environmental conditions that lead to the emergence of atomic cooling haloes which are both metal-free and star-free. 
\begin{figure*} \label{Fig:NumberDCBHs}
\centering
\begin{minipage}{175mm}      \begin{center} 
\centerline{
\includegraphics[width=0.525\textwidth]{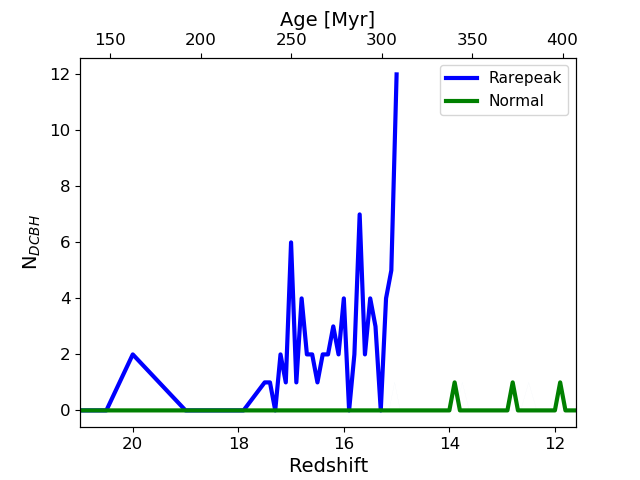}
\includegraphics[width=0.525\textwidth]{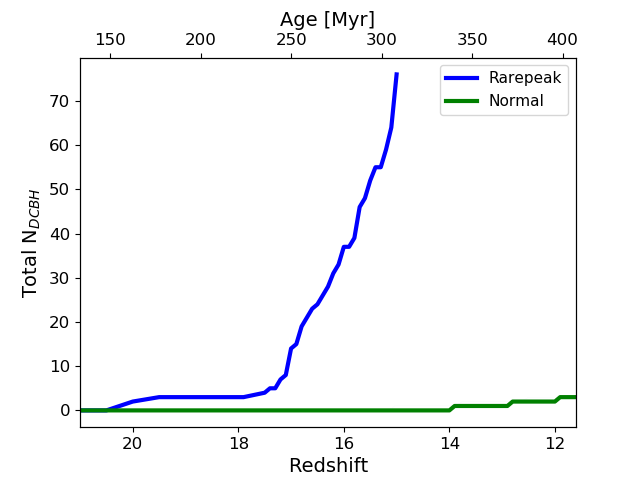}}
\caption{\textit{Left Panel}: The number of DCBH candidate haloes found at each redshift in each region. \textit{Right Panel}: The total number of DCBH candidate haloes found as a function of redshift.  The \rarepeak region (blue line) has formed a total of 76 candidate DCBH haloes. The \normal region (green line) has formed a total of 3 DCBH candidate haloes. The running total is the total number of 
DCBH candidate haloes formed over the entire simulation once duplicates are excluded and accounting for a DCBH candidate halo becoming subsequently polluted. For completeness the age of the universe at that time is included at the top of each figure. }
\end{center} \end{minipage}

\end{figure*}
\section{Renaissance Simulation Suite} \label{Sec:Renaissance}
The Renaissance simulations were carried out on the Blue Waters supercomputer facility using the adaptive mesh refinement code \enzo\citep{Enzo_2014}\footnote{https://enzo-project.org/}. \enzo has been extensively used to study the formation of structure in the early universe \citep{Abel_2002, OShea_2005b, Turk_2012, Wise_2012b, Wise_2014, Regan_2015, Regan_2017}. In particular \enzo includes a ray tracing scheme to follow the propagation of radiation from star formation and black hole formation \citep{WiseAbel_2011} as well as a detailed multi-species chemistry model that tracks the formation and evolution of nine species \citep{Anninos_1997, Abel_1997}. In particular the photo-dissociation of \molH is followed, which is a critical ingredient for determining the formation of the first metal-free stars \citep{Abel_2000}. 

The datasets used in this study were originally derived from a simulation of the universe in a 28.4 \mpch on the side box using the WMAP7 best fit cosmology. Initial conditions were generated using MUSIC \citep{Hahn_2011} at z = 99. A low resolution simulation was run until z = 6 in order to identify three different regions for re-simulation \citep{Chen_2014}. The volume was then smoothed
on a physical scale of 5 comoving Mpc, and regions of high
($\langle\delta\rangle \equiv \langle\rho/\rangle(\Omega_M \rho_C) - 1 \simeq 0.68$), average ($\langle\delta\rangle \sim 0.09)$), and low ($\langle\delta\rangle \simeq -0.26)$)
 mean density were chosen for re-simulation. 
These sub-volumes were then refered to as the \rarepeak region, the \normal region  and the \void region. The \rarepeak region has a comoving volume of 133.6 Mpc$^3$, the \normal region and the \void regions have comoving volumes of 220.5 Mpc$^3$. Each region was then re-simulated with an effective initial resolution of $4096^3$ grid cells and particles within these sub-volumes of the larger initial simulation. This gives a maximum dark matter particle mass resolution of $2.9 \times 10^4$ \msolarc. For the re-simulations of the \voidc, \normal and \rarepeak regions further refinement was allowed throughout the sub-volumes up to a maximum refinement level of 12, which corresponded to 19 pc comoving spatial resolution. Given that the regions focus on different over-densities each region was evolved forward in time to different epochs. The \rarepeak region, being the most over-dense and hence the most computationally demanding at earlier times, was run until z = 15. The \normal region ran until z = 11.6, and the \void region ran until z = 8. In all of the regions the halo mass function was very well resolved down to M$_{halo} \sim 2 \times 10^6$ \msolarc. The \rarepeak regions contained 822 galaxies with masses larger than $10^9$ \msolar at z = 15, the \normal region contained 758 such galaxies at z = 11.6, while the \void region contained 458 such galaxies at z = 8. 

As noted already in \S \ref{Sec:Introduction}, in W19 we examined two metal-free and star-free haloes from the 
\rarepeak simulation. Only the z=15 dataset was used. In this work we examine all of the datasets available
from the \voidc, \normal and \rarepeak regions to get a larger sample of the emergence of DCBH haloes across
all three simulations and across all redshift outputs. In the next section we examine both the number density of DCBH across time and also the environmental conditions which lead to their appearance. 
\section{Results}
We investigate here the emergence of DCBH candidate haloes in the Renaissance simulations. We first investigate the absolute number of DCBH candidate haloes which form in each of the three simulation regions. We then examine in more detail the physical conditions which allow their emergence. 

\subsection{The abundance of DCBH candidate haloes}
In the left panel of Figure \ref{Fig:NumberDCBHs} we show the absolute number of candidate DCBH haloes in each simulation region over the range of redshift outputs available to us. In the right hand panel we show the running total for the number of candidate DCBH haloes formed over the course of the entire simulation. 
As noted in section \S \ref{Sec:Renaissance} the \rarepeak simulation runs to z = 15, the \normal simulation runs to z = 11.6 and the \void simulations runs to z = 8. At each redshift snapshot we calculate the number of \textit{metal-free, atomic cooling haloes which contain no stars}. The number of these DCBH candidate haloes, N$_{DCBH}$, versus redshift is captured in the left hand panel of Figure \ref{Fig:NumberDCBHs}. The \rarepeak simulation (blue line) contains the largest absolute number of DCBH 
candidate haloes. At the final output time (z = 15) there are 12 candidate DCBH haloes in the \rarepeak volume. This compares to 0 in the \normal volume at z = 11.6. However, there are candidates detected in the \normal region at other outputs as we can see. No candidates are detected in the void region at any redshift output and hence we do not explore the \void region any further in this work.

We can see that the number of DCBH candidate haloes fluctuates over time although overall the trend is that there is an increase in the number of the DCBH candidate haloes per unit redshift. The increase is more 
prominently seen in the right hand panel of Figure \ref{Fig:NumberDCBHs}. The running total for the number of DCBH candidate haloes increases rapidly and by z = 15 the \rarepeak simulation has hosted 76 DCBH halo candidates while the \normal region has hosted 3 DCBH halo candidates. The cumulative total accounts for the fact that a previous DCBH candidate halo can become polluted and hence no longer matches the criteria even though it may now host a DCBH\footnote{Renaissance has no subgrid model for DCBH formation and so DCBH is not recorded as haloes assemble.}. In contrast the left hand panel is a pure snapshot at that time and has no memory of the history of haloes. 
%
\begin{figure*}
\centering
\begin{minipage}{175mm}      \begin{center}
\centerline{
    \hspace*{0.35cm}\includegraphics[width=10.5cm, height=10cm]{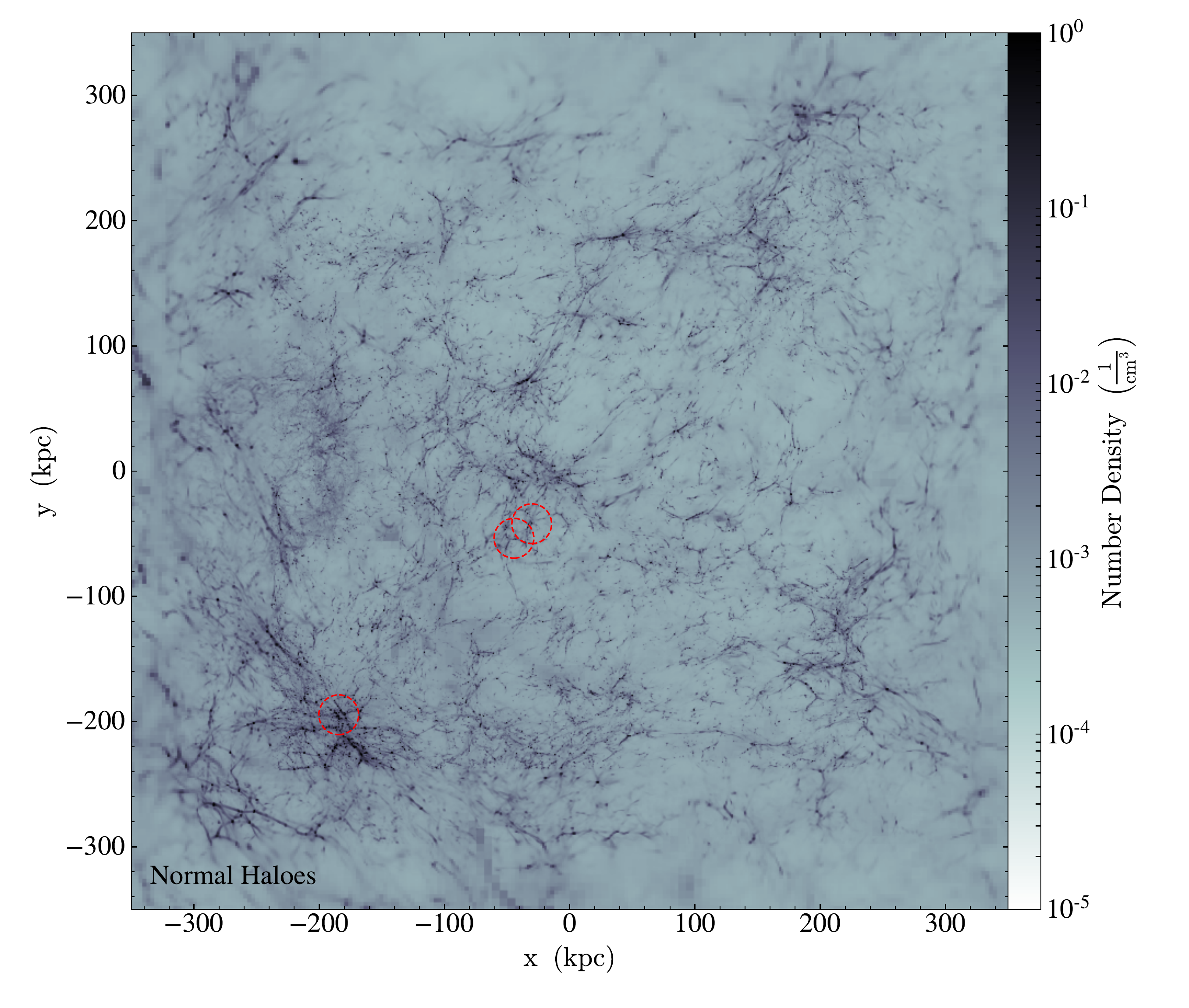}
     \includegraphics[width=10.5cm, height=10cm]{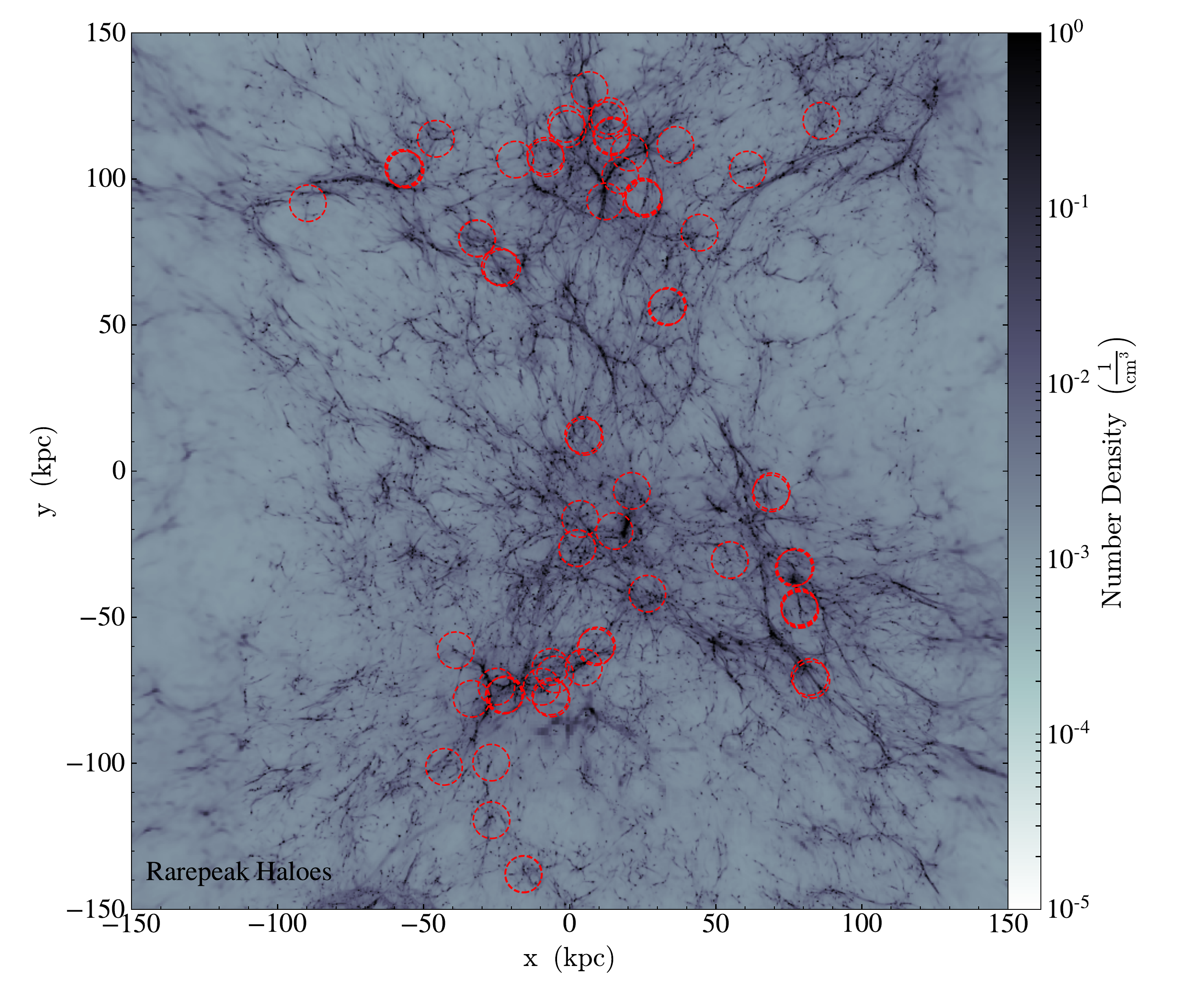}}
\caption{Left Panel: Projection of the \normal simulation volume with dashed red circles identifying the location of all 3 DCBH halo candidates across all redshift outputs.  Right Panel: Projection of the \rarepeak simulation volume with dashed red circles identifying the location of all 76 
DCBH candidates across all redshift outputs. The \rarepeak projection is made at z = 15 and the \normal projection is made at z = 11.6 although the DCBH candidate haloes may have formed at a different epoch.}
\label{Fig:Projections}
\end{center} \end{minipage}
\end{figure*}
In Figure \ref{Fig:Projections} we plot the location of each of the distinct DCBH candidate haloes on top of
a projection of the number density of the \rarepeak region and of the \normal region. In each case the 
projection is made at the final redshift output (\textit{Rarepeak}, z=15; \textit{Normal}, z=11.6). The dashed red circles which denote the halo location are from across all redshift outputs and hence should be seen as approximate locations. Nonetheless, what is immediately obvious is that the emergence of DCBH candidate haloes is a ubiquitous feature of high density regions. The number of haloes in the \normal region is significantly reduced compared to the \rarepeak region. The reason behind the much larger number of DCBH candidates in the 
\rarepeak region compared to the \normal region is multifaceted, depending on the growth of structure, the mean density of the inter-galactic medium in that region and the flux of LW radiation.

The number of galaxies above some given minimum mass M$_{min}$(z) in a redshift bin of width $dz$ and solid angle $d\Omega$ can be defined using the Press-Schechter formalism \citep{PressSchecter_1974}. 
\begin{equation}
    {{dM} \over {d\Omega dz}}(z) = {{dV} \over {d\Omega dz}}(z) \int^{\inf}_{M_{min}(z)} dM {dn \over dM} (M, z)
\end{equation}
where $dV / d\Omega dz$ is the cosmological comoving volume element at a given redshift and $(dn / dM)dM$ is the 
comoving halo number density as a function of mass and redshift. The latter quantity was expressed by \cite{Jenkins_2001} as 
\begin{equation}
\begin{aligned}
{dn \over dM}(M, z) = {} &  -0.315 {\rho_o \over M} {1 \over \sigma_M} {d\sigma_M \over dM} \times \\
                         & \exp ({-|0.61 - log(D(z) \sigma_M)|^{3.8})})
\end{aligned}
\end{equation}
where $\sigma_M$ is the RMS density fluctuation, computed on mass scale $M$ from the $z = 0$ linear power spectrum \citep{Eisenstein_Hu_1999}; $\rho_0$ is the mean matter density of the universe, defined as $\rho_0 = \Omega_M*\rho_c$ (with $\rho_c$ being the cosmological critical density, defined as $\rho_c = 3H_0^2/8 \pi G$) and $D(z)$ is the linear growth function (see, e.g. \cite{Hallman_2007} for details). Taking this together we find that $dn / dM $ scales approximately as $\rho \sigma_{M}^{3.8}$. 

The higher mean density and higher 
$\sigma_M$ in the \rarepeak compared to the \normal region is therefore consistent with previous findings showing that there are  approximately 3 - 4 times more haloes, per unit redshift, in the \rarepeak region 
\citep{Xu_2013, OShea_2015}. Not only this but the higher mean densities in the \rarepeak region leads to 
a smaller volume filling fraction of metal enrichement in the \rarepeak region compared to the \normal region. 
Taking supernova blastwave calculations alone leads to a volume filling fraction of 0.7 in the \rarepeak 
relative to the \normal region. Finally, the flux of LW is also much higher in the \rarepeak region as there
are more haloes producing more stars per unit volume compared to the \normal region (see e.g. \cite{Xu_2013}). The combination of these three factors leads to significantly more DCBH candidate haloes in the \rarepeak region. Over the time interval that the Renaissance simulations run for this leads to a ratio of 76 DCBH candidates in the \rarepeak region compared to just 3 in the \normal region. 
\begin{figure*}
\centering
\begin{minipage}{175mm}      \begin{center}
\centerline{
    \includegraphics[width=9.5cm, height=8cm]{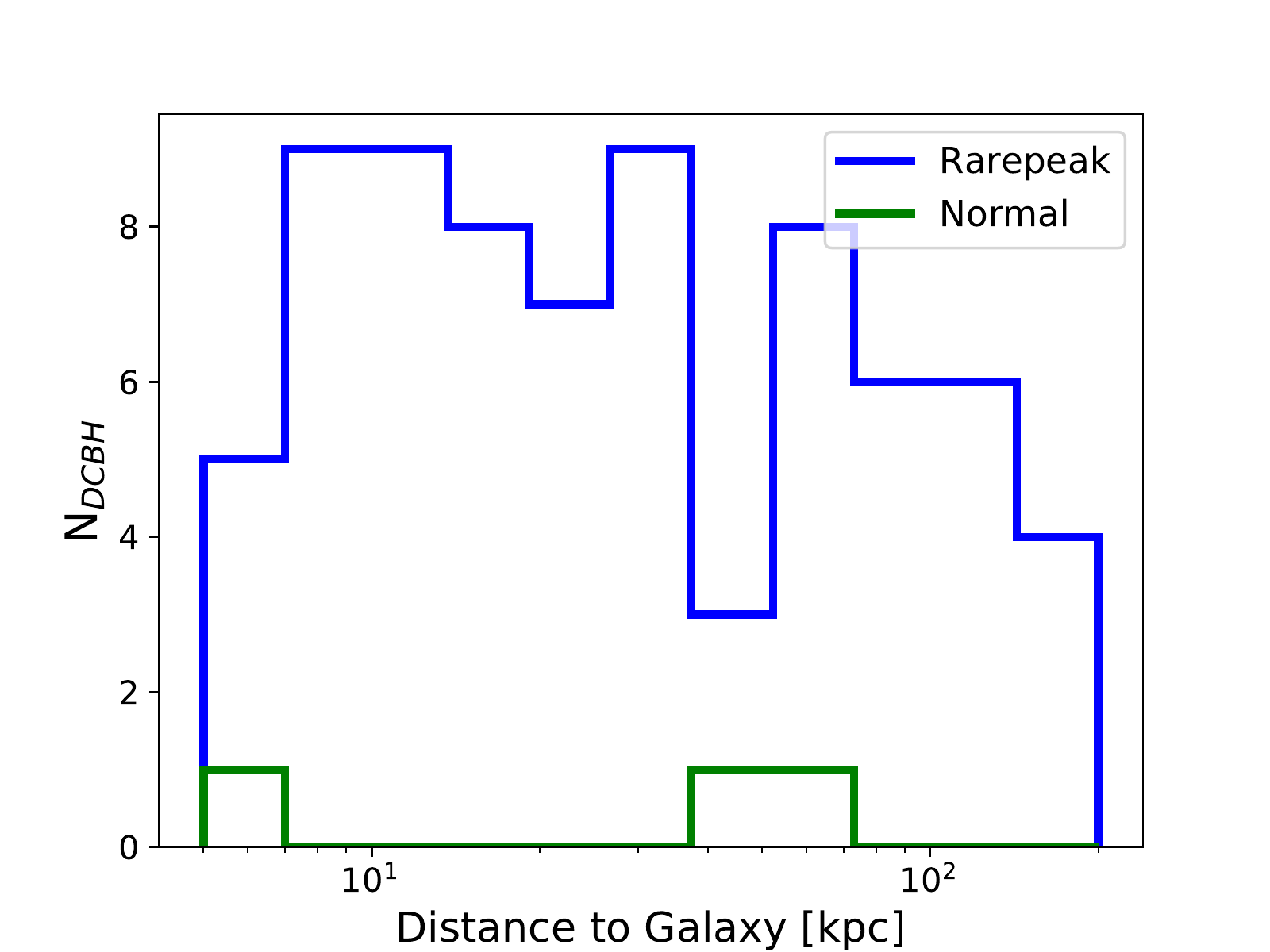}
     \includegraphics[width=9.5cm, height=8cm]{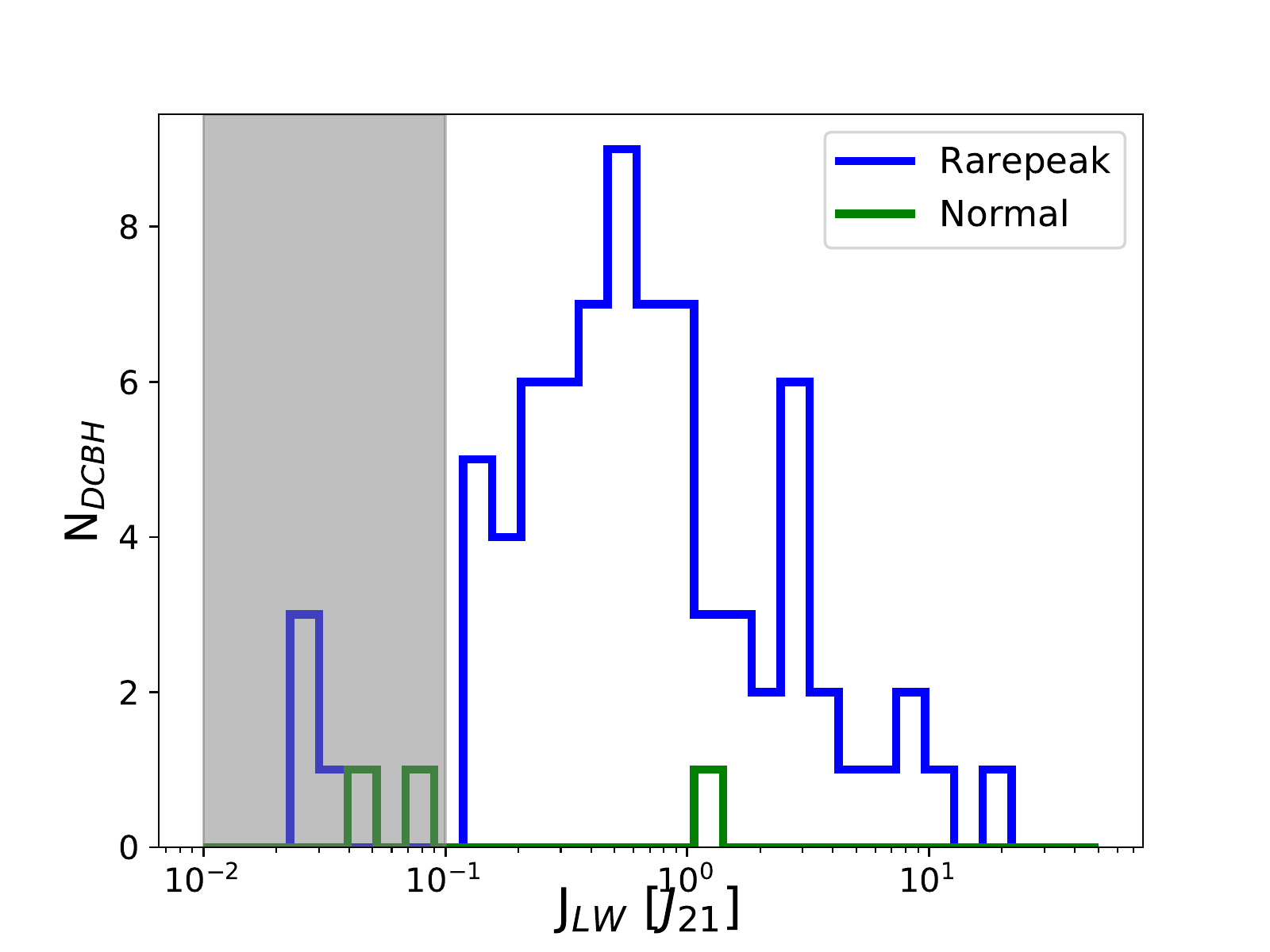}}
\caption{Left Panel: The distance from each candidate DCBH halo to the nearest massive galaxy (defined as
the closest star forming halo, see text for more details) for each 
region. Right Panel: The value of the LW background, in units of J$_{21}$, felt at the centre of each DCBH 
candidate. For the majority of DCBH haloes the value of LW radiation it is exposed to is within an order of magnitude of the background level at that redshift. Only a small number of DCBH candidate haloes experience radiation levels more than one order of magnitude higher than the background level. The grey vertical band indicates the approximate level of background LW radiation expected at z = 15 \citep{Ahn_2009, Xu_2013}.}
\label{Fig:DistJ21}
\end{center} \end{minipage}
\end{figure*}

\subsection{The physical conditions required for DCBH candidate halo formation}
In Figure \ref{Fig:DistJ21} we plot the distance from each DCBH candidate halo to the nearest massive 
galaxy and we also plot the level of LW radiation that each candidate halo is exposed to. In the left hand panel of Figure \ref{Fig:DistJ21} the distance\footnote{All distances discussed are in physical units unless explicitly stated otherwise} to the nearest massive galaxy (defined below) is calculated by examining every halo in a sphere of radius 1 Mpc around the DCBH candidate halo. The stellar mass in each halo is then normalised by the square of the distance between that halo and the candidate halo. This normalisation accounts for the $r^{2}$ drop off in radiation intensity with distance. The galaxy with the largest normalised stellar mass is then used as the nearest massive galaxy. In the \rarepeak 
simulation most galaxies lie at least 10 kpc away but the spread is quite even up to nearly 100 kpc at which point it starts to decline. In the \normal simulation, which only has 3 candidates, the nearby galaxies lie 
approximately 5 kpc and 50 kpc (in two of the cases) away. What this tells us is that close proximity to nearby 
star-forming galaxies is not (directly) correlated with forming DCBH candidate haloes.
In the right hand panel we investigate the level of LW radiation that each candidate halo is exposed to 
at the associated redshift output. In this case the results are somewhat more defined. For the \rarepeak 
region the values of J$_{LW}$ are between 0.01 and 10 J$_{21}$ while for the 
\normal simulation the values are between approximately 0.1 and 1 J$_{21}$, albeit for significantly fewer DCBH candidate haloes. The values for the LW radiation field, in the \rarepeak region, are approximately an order of magnitude higher than the expected mean radiation field at this redshift of J$_{LW} = 10^{-2} - 10^{-1}$ J$_{21}$  \citep{Ahn_2009, Xu_2013} - marked by the shaded region in Figure \ref{Fig:DistJ21}. The reason for this is that the \rarepeak region has significantly more galaxies \citep{OShea_2015} compared to the \normal region and the galaxies are also much brighter, especially in the LW band.

The level of LW radiation felt by the vast majority of candidate DCBH haloes is significantly below the level required to fully suppress \molH cooling \citep{Regan_2014b, Latif_2014a, Regan_2016a}, which is typically estimated to be approximately 1000 J$_{21}$. Nonetheless, the haloes do not collapse until after reaching 
the atomic cooling limit. As we found in W19 the impact of rapid halo growth plays a dominant role in the halo assembly history of these haloes, as we now discuss.
\begin{figure*}
\centering
\begin{minipage}{175mm}      \begin{center}
\centerline{
    \includegraphics[width=9.5cm, height=8cm]{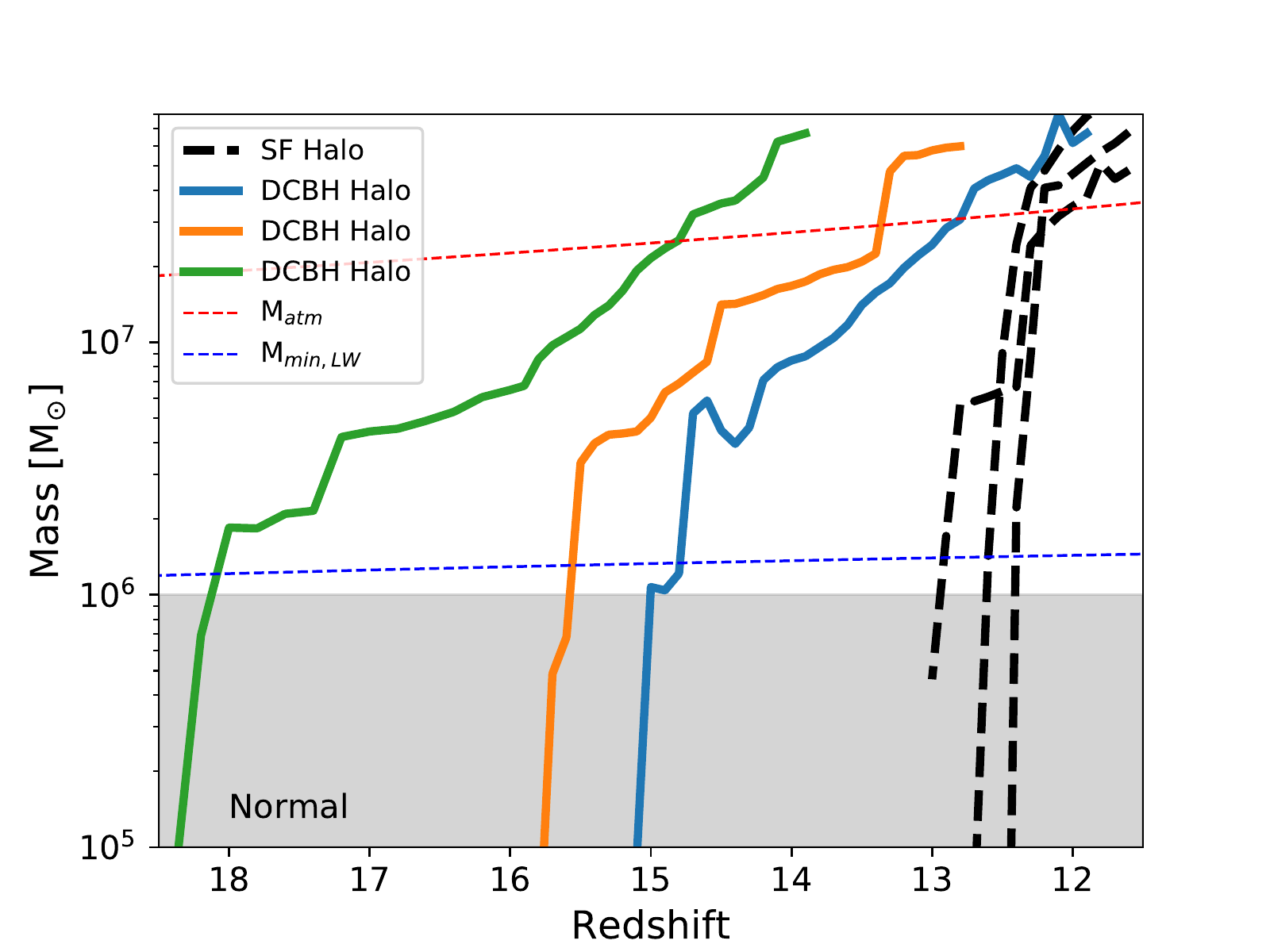}
     \includegraphics[width=9.5cm, height=8cm]{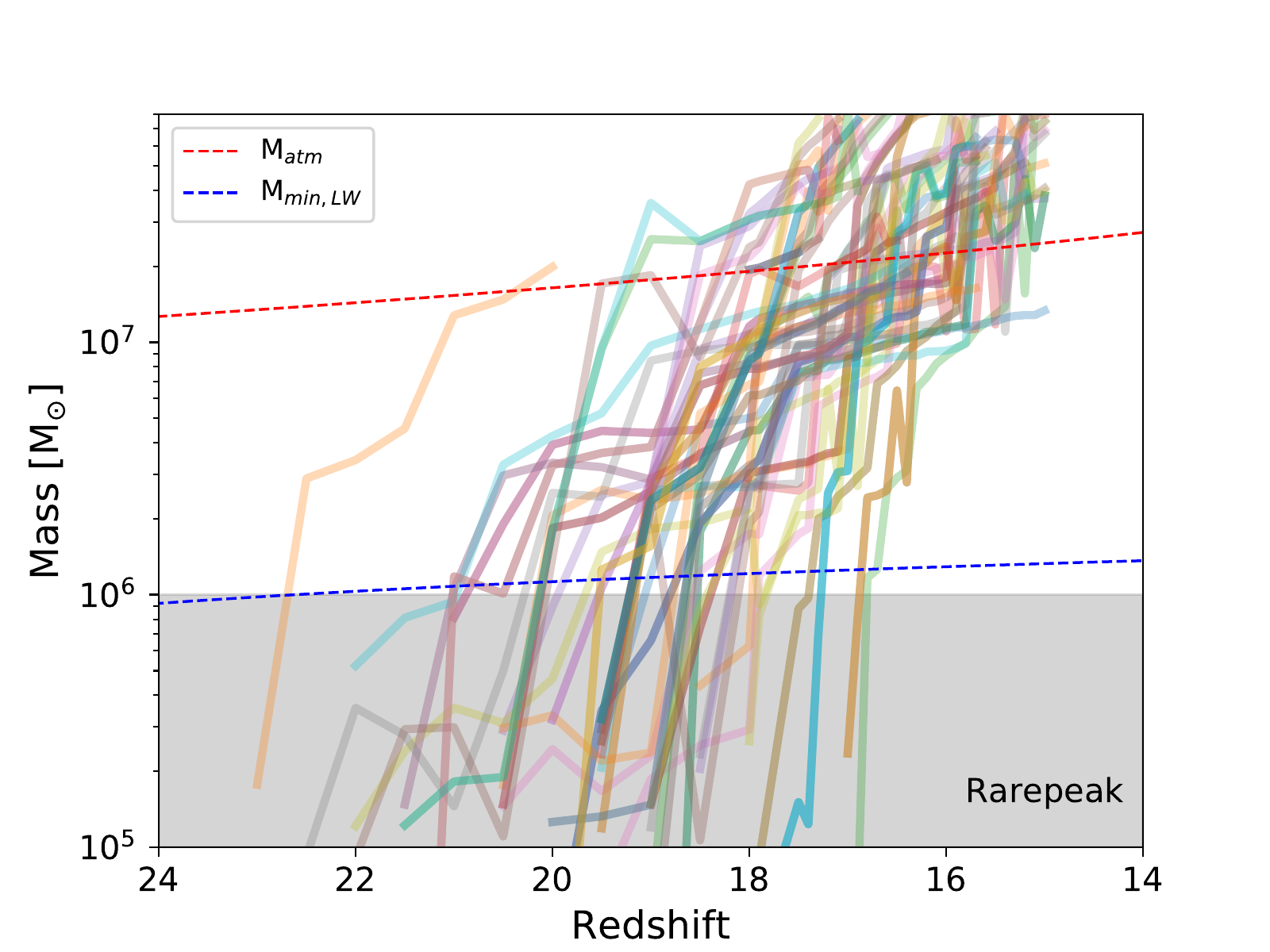}}
\caption{Left Panel: The evolution of the total mass of each DCBH candidate halo in the \normal simulation. Also included (dashed black lines) is the evolution of three rapidly growing star-forming haloes for comparison. The mass
resolution of the Renaissance simulations is approximately 20,000 \msolar so values below 10$^6$ \msolar should be treated with caution and we therefore set the halo resolution of our analysis at 10$^6$ \msolarc. Right Panel: The evolution of the total mass of each DCBH candidate halo in the \rarepeak simulation.  In the vast majority of cases the halo grows rapidly just prior to reaching the atomic cooling limit.}
\label{Fig:MassGrowth}
\end{center} \end{minipage}
\end{figure*}

In Figure \ref{Fig:MassGrowth} we plot the mass growth of each candidate DCBH halo as a function of redshift. In both panels we plot the mass of the halo versus the redshift. The left panel contains haloes from the \normal simulation while the right hand panel contains haloes from the \rarepeak simulation. The grey region in each panel below $10^6$ \msolar signifies the region below which
the mass resolution of Renaissance becomes insufficient to confidently model haloes. Generally we are able to track haloes below this threshold and into the grey region but below  $10^6$ \msolar results should be treated with caution. The dashed blue line is the limit above which a halo must grow in order to overwhelm the impact of LW radiation, M$_{min,LW}$, \citep{Machacek_2001, OShea_2008, Crosby_2013, Crosby_2016}. The dashed red line is the approximate atomic cooling threshold, M$_{atm}$, at which point cooling due to atomic hydrogen line emission becomes effective\footnote{Both  M$_{min,LW}$ and M$_{atm}$ evolve with redshift although the dependence is weak over the range considered here}. 
Focusing first on the \normal region in the left panel we plot the growth rate of the three DCBH candidate haloes identified in the left panel of Figure \ref{Fig:Projections}. The DCBH candidate haloes are rapid growers but are not necessarily the fastest growing haloes in the \normal region. To emphasise this comparison we also plot the growth of three rapidly growing haloes which contain stars. We select the three star forming haloes from the final output of the \normal region but haloes at other redshifts do of course exist which are rapidly growing and contain stars. 
In this case we see that haloes with high dM/dz (i.e. the mass as a function of redshift) values can be star-free or star-forming and hence having a high dM/dz does not necessarily discriminate between DCBH halo candidates by itself. Rapidly growing
haloes can become metal-enriched through external enrichment processes. The enrichment allows the halo
interior to cool and to form stars even in the presence of dynamical heating. Therefore, any semi-analytical model or sub-grid prescription which uses dM/dz alone as a predictor for DCBH candidates will inevitably 
overestimate the number of candidates. 

The right hand panel of Figure \ref{Fig:MassGrowth} shows the growth of DCBH candidate haloes from the \rarepeak simulation. There is a much larger number of DCBH candidate haloes in the \rarepeak region 
compared to the \normal region and hence only the DCBH candidate haloes are included in this plot. Again we see strong evidence of rapid assembly. All of the haloes show evidence of rapid growth between the LW threshold and the atomic cooling limit, which is able to suppress star formation in all of these haloes. The dynamics of each halo are somewhat unique, with some haloes experiencing major mergers that lead to bursts of dynamical heating while others experience more steady but nonetheless rapid growth. Furthermore, some haloes will be located closer to massive galaxies which expose the haloes to high LW radiation which in-turn impacts the chemo-thermodynamical characteristics of the halo in question. We now examine the roles that metallicity, rapid growth and radiation all play in the assembly of a DCBH candidate halo in more detail. 

\begin{figure*}
\centering
\begin{minipage}{175mm}      \begin{center}
\centerline{
    \includegraphics[width=9.5cm, height=8cm]{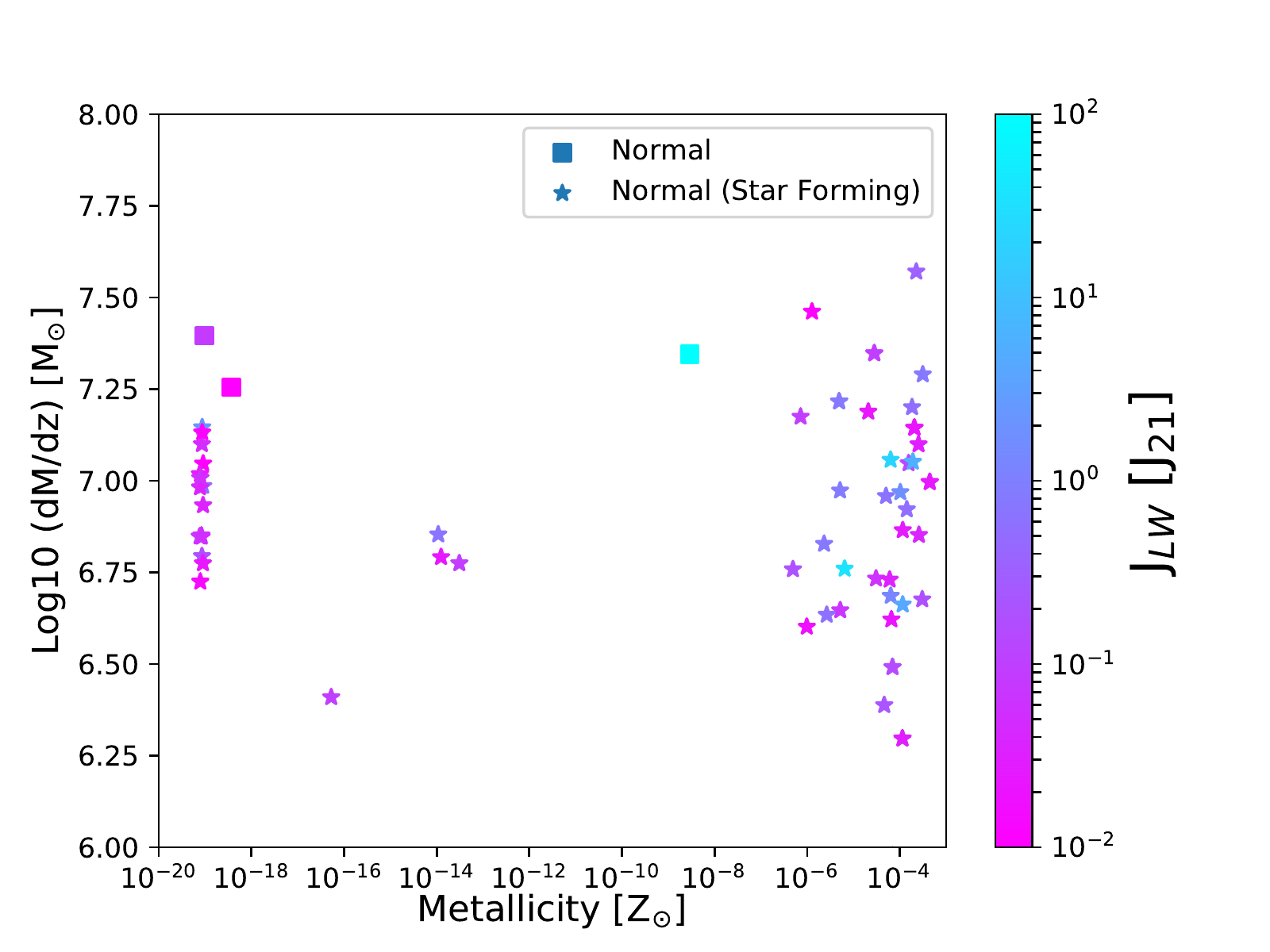}
    \includegraphics[width=9.5cm, height=8cm]{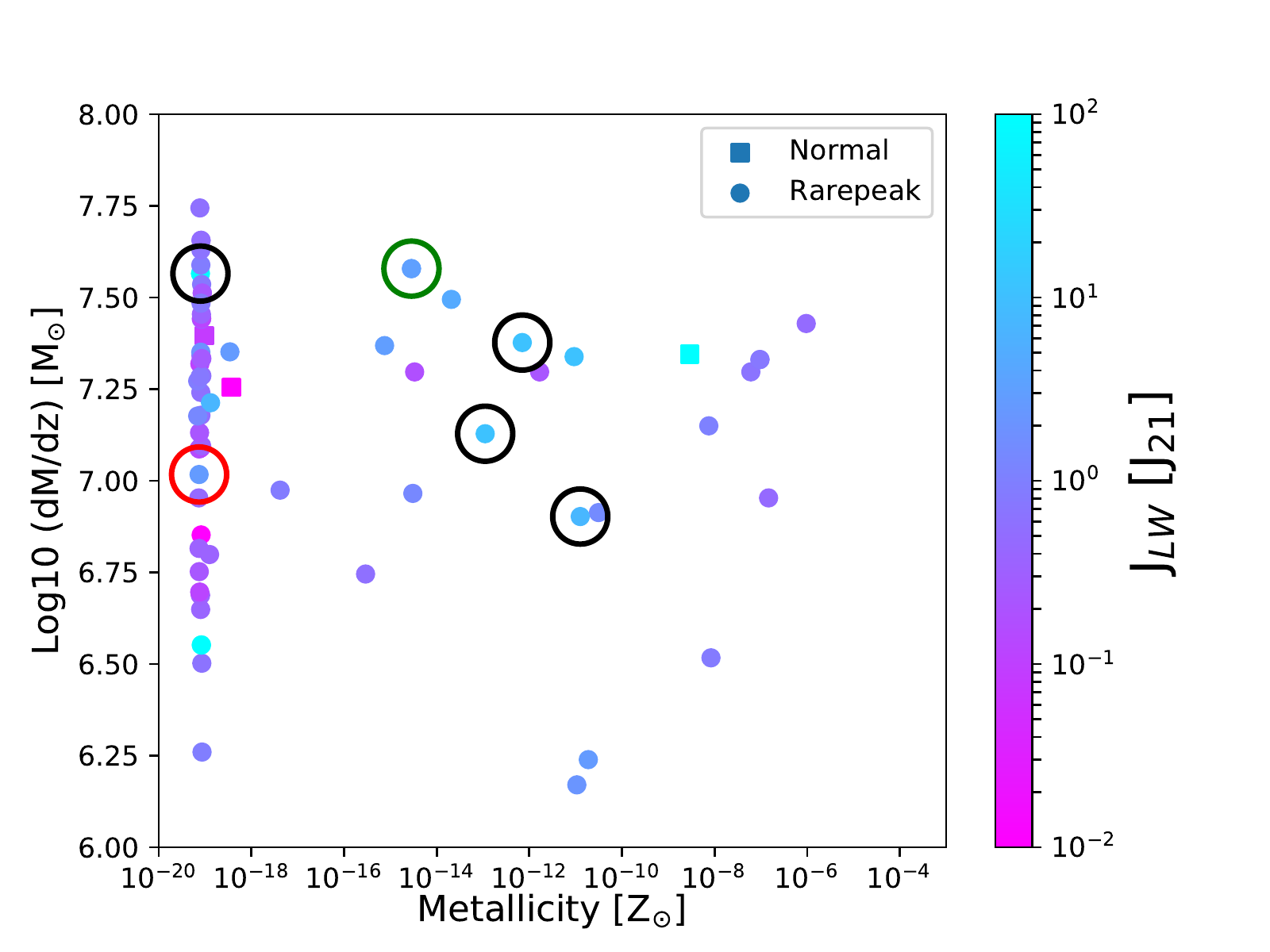}}
\caption{Left Panel: Phase space diagram showing the maximum rate of growth (dM/dz) of the DCBH candidate haloes in the \normal region (squares). Also included is the growth rate of a large sample of star-forming haloes for comparison.  It should be noted that while the DCBH candidate haloes are among the most rapidly growing haloes, star forming haloes can grow more rapidly. The colour of the squares, stars and circles are weighted by the LW radiation to which that halo is exposed prior to the onset of star-formation. Right Panel: Similar plot for the \rarepeak simulation. The growth rate, dM/dz, for DCBH candidates in the \rarepeak simulation are shown as circles again coloured by the level of LW radiation to which they are exposed. The 
DCBH candidates haloes from the \normal simulations are also plotted for direct comparison. Black outer circles are used to identify four DCBH candidates which collapse completely isothermally at T = 8000 K. The DCBH candidate halo marked with a red outer circle is the MMHalo from W19 while the green outer circle is the LWHalo from W19.}
\label{Fig:Scatter}
\end{center} \end{minipage}
\end{figure*}

\subsection{Radiation, Metallicity \& Rapid Growth all play a role}

In Figure \ref{Fig:Scatter} we examine quantitatively the dM/dz values from haloes in both the \normal and \rarepeak regions. We compare in a 3D representation the average dM/dz, \JLW \ and metallicity of 
each of the DCBH candidate haloes as well as a subset of star-forming haloes from the \normal region.
In the left hand panel of Figure \ref{Fig:Scatter} we focus on the \normal region. The phase diagram 
shows the average growth rate, dM/dz, as a function of halo metallicity. Each symbol is coloured by the 
level of LW radiation the halo is exposed to. We plot the dM/dz, metallicity and \JLW \ values of both DCBH candidate haloes (squares) and star-forming (stars) haloes. The dM/dz value is calculated by determining the time taken for a halo to grow from $5 \times 10^6$ \msolar up to the atomic cooling limit ($\sim 3 \times 10^7$ \msolarc). This measures the mean rate at which mass is accumulated by the halo once it crosses the LW threshold (blue line in Figure \ref{Fig:MassGrowth}) and up to the point it reaches the atomic cooling limit (red line in Figure \ref{Fig:MassGrowth}). Both the J$_{LW}$ value and the metallicity are calculated by taking the final value of J$_{LW}$ and metallicity respectively before star formation occurs (star formation leads to additional internal LW radiation and metal enrichment which we cannot disentangle from external effects). The three DCBH candidate haloes have among the highest dM/dz values, which goes some way to explaining why these haloes were able to suppress star formation. The dynamical heating impact of rapid growth is given by
\begin{equation}
    \Gamma_{\rm{dyn}} = \alpha M_{\rm{halo}}^{-1/3} {{k_b} \over {\gamma -1}} {{ dM_{\rm{halo}} \over dt}}
\end{equation}
where $\Gamma_{\rm{dyn}}$ is the dynamical heating rate, $ M_{\rm{halo}}$ is the halo total mass and $\alpha$
is a coefficient relating the virial mass and temperature of the halo \citep{Barkana_2001}. Two of the haloes are completely metal-free while one of the haloes is experiencing some slight external metal enrichment ($\sim 2.88 \times 10^{-9}$ 
\zsolarc). However, it is also clear that there are star-forming haloes growing more rapidly than the star-free haloes. This is not surprising. In the case of the halo in the top right of the left panel this halo became metal enriched early in the halo assembly process. The halo formed a PopIII star but the halo continued to assemble rapidly. In this case because of the metal enrichment the dynamical heating due to rapid assembly is negated completely. Therefore, only haloes which remain metal-free \textit{and} grow rapidly can remain star-free. 

In the right hand panel of Figure \ref{Fig:Scatter} we plot the same phase plot for the DCBH candidate haloes (circles) in the \rarepeak simulation. Given the large number of DCBH candidate haloes in the \rarepeak region we do not include star-forming haloes from the \rarepeak region in this plot. We do, however, include the DCBH candidate haloes (squares) from the \normal region for direct comparison. For these DCBH candidate haloes there is a wide variation in Log$_{10}$ (dM/dz) with values as low as 6.3 and as high as 7.75. 
Naively it would be expected that the haloes with low dM/dz values and moderate to low \JLW values would
form stars. However, inspection of individual haloes reveals bursts of rapid assembly which can result in 
the suppression of \molH for at least a sound crossing time (see also W19). The average value of dM/dz, as plotted here, fails to detect the bursts which can suppress star formation and in many cases those with low average dM/dz values have a strong burst of dynamical heating not easily captured by an average value. We will return to this point and the impact this can have on deriving a semi-analytic prescription in \S \ref{Sec:Discussion}.

\begin{figure*}
\centering
\begin{minipage}{175mm}      \begin{center}
\centerline{
    \includegraphics[width=18.0cm, height=12cm]{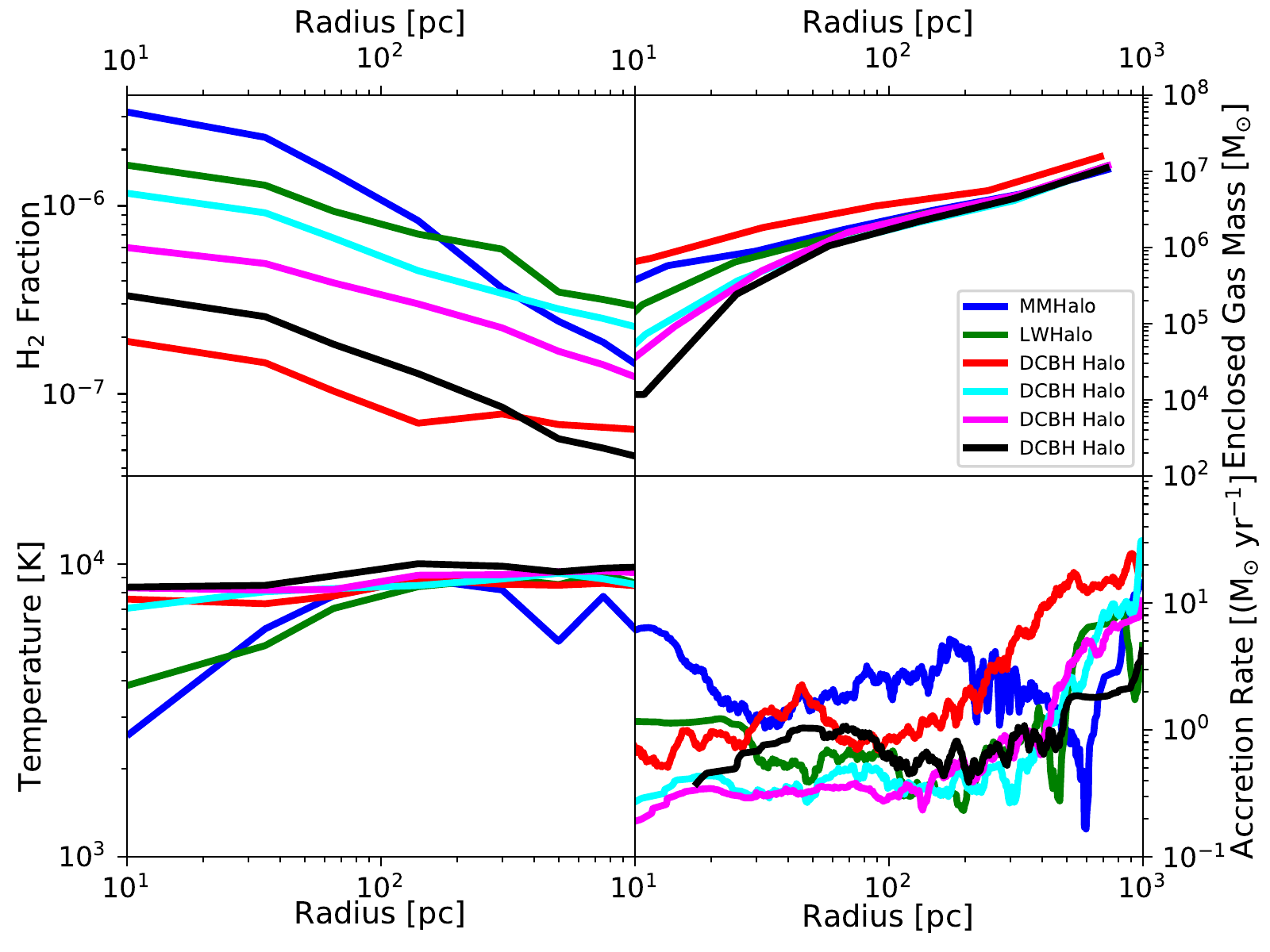}}
\caption{In each of the four panels in this figure we compare the six DCBH haloes identified in the right hand panel of Figure \ref{Fig:Scatter}. Four of the DCBH candidate haloes are collapsing isothermally while the 
MMHalo (blue line) and the LWHalo (green line) show strong evidence of non-isothermal collapse. In the bottom left hand panel we plot the temperature against radius illustrating the isothermality of the four selected DCBH candidate haloes. The MMHalo and the LWHalo clearly start to cool in the halo centre. This cooling can be directly attributed to a higher \molH fraction for the MMHalo and the LWHalo as seen in the top left panel. The enclosed mass for each candidate halo varies inside approximately 30 pc for each halo with an average enclosed mass of $10^5$ \msolar inside 20 pc. In the bottom right panel we show the instantaneous accretion rate for each DCBH candidate halo. All of the haloes show accretion rates greater than 0.1 \msolarc/yr across several decades in radius and continuing into the core of the halo. }
\label{Fig:Rarepeak_Profiles}
\end{center} \end{minipage}
\end{figure*}

In the right hand panel we identify six haloes with circles. Four are marked with black circles. These are haloes that we have found show an isothermal collapse up to the maximum resolution of the Renaissance simulations ($\sim 1$ pc) and are showing no signs of \molH cooling in the core of the halo. Each of the isothermal haloes that we identify here are typically within a few kiloparsecs of a star forming atomic cooling halo but the candidate halo has not yet become either significantly metal enriched or photo-evaporated. Nonetheless, the nearby massive galaxies provided a much higher than average (average \JLW $\sim 1$ J$_{21}$) \JLW \ value. This scenario is similar to the scenario explored by \cite{Dijkstra_2014}. We also identify in red the most massive halo in the \rarepeak simulation at z = 15 and the most irradiated halo (green circle) in the simulation at z = 15. The most massive and most irradiated halo were previously identified in W19 and 
investigated in detail. \\
\indent In Figure \ref{Fig:Rarepeak_Profiles} we show the radial profiles of a number of physical quantities for each of the haloes identified by the circles. The blue line is the most massive halo (MMHalo) and the green line (LWHalo) is the
most irradiated halo. The other haloes are those which show well defined isothermal collapse profiles. Both the MMHalo and the LWHalo show clear cooling towards the molecular cooling track (bottom left panel). Each of the other haloes have temperatures greater than 8000 K all the way in to the centre of the halo and so remain on the cooling atomic cooling track. In the top left panel we see that both the MMHalo and the LWHalo have higher \molH fractions as expected. All the haloes increase their \molH as the density increases towards the centre of the halo. In the case of the isothermally collapsing haloes the fraction remains low enough so that cooling remains dominated by atomic cooling. In the top right panel we plot the enclosed gas mass as a function of radius and in the bottom right panel the instantaneous accretion rate as a function of radius. The accretion rates for each of the haloes are extremely high, with accretion rates above 0.1 \msolar per year at all radii. Accretion rates greater than approximately 0.01 \msolarc/yr are thought be required for supermassive star formation \citep[e.g.][]{Sakurai_2016, Schleicher_2013}. The MMHalo and the LWHalo cool towards the centre of the halo, 
meaning that fragmentation into a dense cluster of PopIII stars becomes more likely in those cases.
The reason that the MMHalo and the LWHalo cool towards the centre is due to their higher \molH
  fractions compared to the other four haloes. As can be seen in Figure \ref{Fig:Scatter} each of the four
  selected haloes have systematically higher LW radiation values impinging onto them resulting in lower \molH
fractions.
In addition, for the cases where the collapse remains isothermal the degree of fragmentation can be suppressed, with more
massive objects likely to form in that case \citep{Regan_2018a, Regan_2018b}.

\begin{figure*}
\centering
\begin{minipage}{175mm}      \begin{center}
\centerline{
    \includegraphics[width=18.0cm, height=12cm]{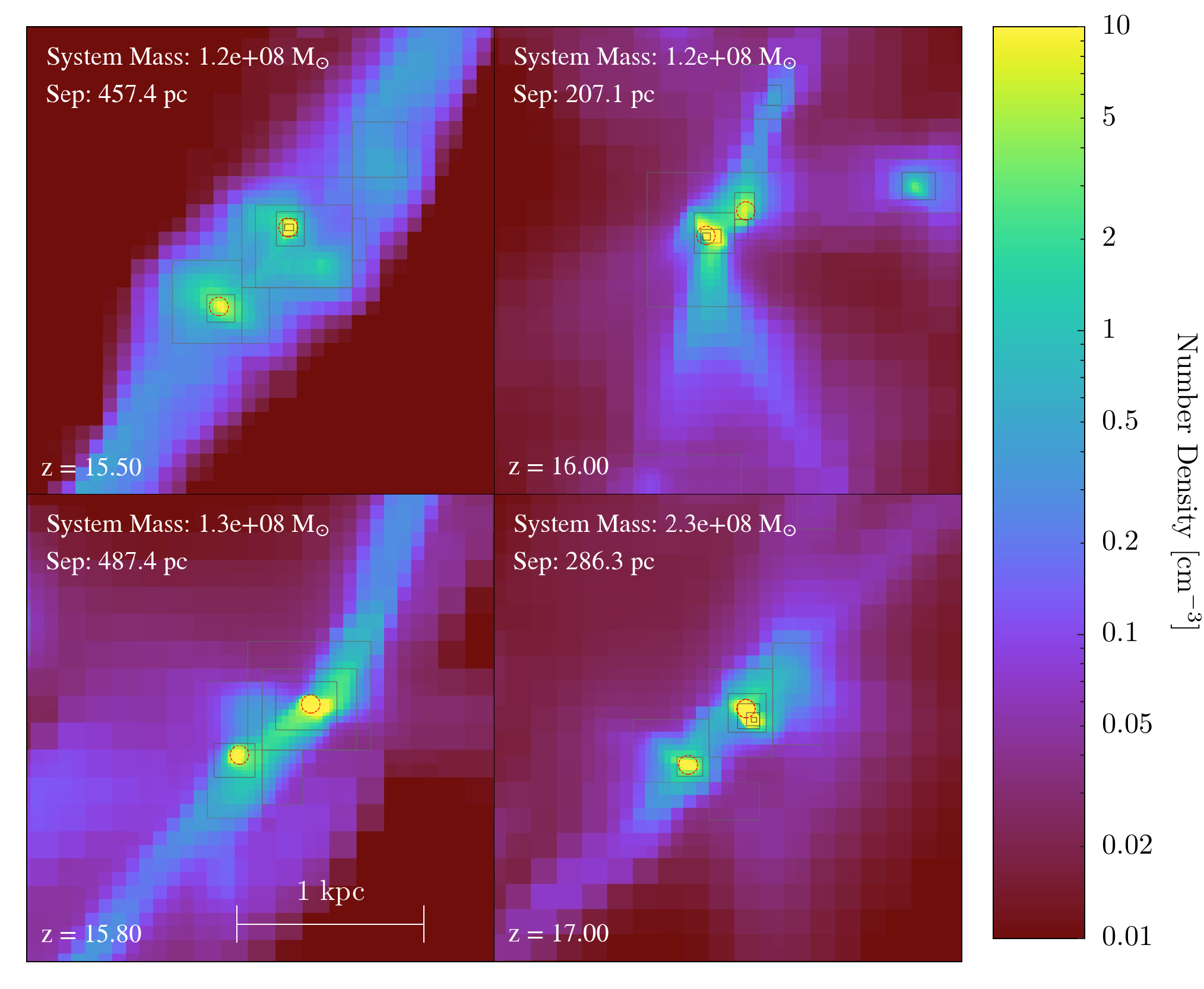}}
\caption{Visualisations of four of the synchronised haloes found in the \rarepeak region. Each member of the synchronised pair is an atomic cooling halo on the cusp of star formation. Typical separations between haloes are between 200 pc and 500 pc at these outputs. The red circles in each panel mark the central core of each halo. The radius of each circle is approximately 10\% of the virial radius. The virial radius of each individual DCBH candidate haloes overlaps with its synchronised partner halo. Only the system mass is shown in each panel since the haloes are subhaloes of each other. 
}
\label{Fig:SyncHaloes}
\end{center} \end{minipage}
\end{figure*}
\subsection{Synchronised Haloes} \label{Sec:Synchronised}
Synchronised haloes have been invoked as a means of generating a sufficiently high LW radiation flux to allow
total suppression of \molH in the core of an atomic cooling halo \citep{Dijkstra_2008, Visbal_2014b, Regan_2017}. The scenario supposes that two pristine progenitor atomic cooling haloes cross the atomic cooling threshold nearly simultaneously. The suppression of star formation in both haloes as they assemble eliminates the possibility of either metal enrichment or photo-evaporation from one halo to the other. The first halo to cross the atomic cooling threshold suffers catastrophic cooling due to neutral hydrogen line emission cooling and begins to collapse and form stars. The LW radiation from Halo1 irradiates Halo2, thus suppressing \molH in Halo2 and allowing for the formation of a DCBH. We search the \rarepeak region for synchronised pairs matching the above criteria.

We look for pairs of ACHs which remain pristine and devoid of star formation and are separated from each other by less than 1 kpc, but are also at a separation of greater than 150 pc as they cross the atomic cooling threshold. We note that this is likely somewhat optimistic given the region of synchronisation is expected to be between approximately 150 pc and 350 pc for haloes 
of this size \citep{Regan_2017}. Within the \rarepeak region we find of total of 5 pairs of pristine ACHs that fulfill the basic criteria. In Figure \ref{Fig:SyncHaloes} we show a visualisation of four of the
five haloes which are candidates for synchronised haloes. In each case the haloes are 
separated by distances between approximately 200 pc and 500 pc at the time of crossing the atomic cooling threshold. In all cases the haloes are still devoid of star formation but at least one of the haloes in the 
pair forms stars before the next data output. The total mass of the two atomic cooling haloes in each case
is above $10^{8}$ \msolarc. Given the proximity of the two haloes at this point it is difficult to estimate the mass of each halo individually.

\cite{Visbal_2014} examined the formation of DCBH from synchronised haloes and estimated their abundances both analytically and through an n-body simulation. To estimate the 
abundances of synchronised haloes analytically they used the following equation
\begin{equation} \label{Eq:sync}
    {dn_{DCBH} \over dz} \sim {dn_{cool} \over dz} \Big ({dn_{cool} \over dz} \Delta_{z_{sync}} \int^{R.O.R} dr 4\pi r^2 [1 + \eta(r)] f_s(r)  \Big )
\end{equation}
where ${{dn_{cool} \over dz}} $ is the number density of haloes which cross the cooling threshold between $z$ and $z + dz$, $\eta(r)$ is the two point correlation function which describes the enhancement of halo pairs due to clustering and $\Delta_{z_{sync}}$ is the redshift range corresponding to the synchronisation time, $f_s(r)$ is the fraction of haloes that are found at a radius, $r$, when they cross the atomic threshold. \cite{Visbal_2014} used an n-body-only simulation to determine the values required for Equation \ref{Eq:sync}. They predicted 15 synchronised pairs in a 3375 cMpc$^3$ volume. In the \rarepeak region, which has a volume of 133.6 cMpc$^3$, we find 5 synchronised pairs. Given the difference in volume our 
abundance is higher by a factor of approximately 5 compared to that of \cite{Visbal_2014b}. However, the 
\rarepeak region represents an over-density of approximately 1.7 compared to an average region of the universe and \cite{Visbal_2014} also preformed the calculation at a somewhat lower redshift. When this is taken into account our values match those of \cite{Visbal_2014b} quite well. Furthermore, \cite{Visbal_2014b} were unable
to account for metal enrichement in their analysis which may have a led to an over-estimate of the number
density of synchronised halo candidates in that case.

In order to test the feasibility of the synchronised haloes found in this work a zoom-in re-simulation of the region surrounding the synchronised pairs is required which accounts for both normal PopIII star formation, in Halo1, and possible super-massive star formation in Halo2. In order to provide a sufficient flux,
\cite{Regan_2017} predicted that Halo1 must form approximately $10^5$ \msolar of stellar mass in order to generate a significantly strong LW flux to achieve isothermal collapse. However, the DCBH candidate haloes 
found here have already had their ability to form \molH suppressed due to dynamical heating. Therefore, these particular haloes may not require such intense external radiation exposure. Detailed re-simulation of these candidate haloes is now required to quantify the level of LW required in this case. 

\section{Discussion and Conclusions} \label{Sec:Discussion}
We have analysed the Renaissance suite of high resolution simulations of the early Universe with the goal of identifying candidate haloes in which DCBHs can form. In total we found 79 haloes over all redshifts and volumes which have crossed the 
atomic cooling limit and remain both metal-free and star-free. These 79 haloes represent ideal locations in which to form a DCBH as they will shortly undergo rapid collapse due to neutral hydrogen line emission cooling. The nature of the collapse cannot be probed in these simulation as Renaissance has no sub-grid prescription for super-massive star formation and lacks the resolution to accurately track possible fragmentation into a dense stellar cluster of PopIII stars. 

In general the candidate haloes form away from massive galaxies. This allows the candidate haloes to remain free of metal enrichment. In examining the distance that these candidate haloes are from their nearest massive galaxy we find that the DCBH candidate galaxies typically lie between 10 kpc and 100 kpc from the nearest massive galaxy. These massive galaxies provide LW intensities that are approximately one order of magnitude higher than the mean intensity expected at these redshifts \citep{Ahn_2015, Xu_2016}. However, only a small fraction of the candidate haloes are exposed to LW intensities greater than 10 J$_{21}$. We find that the primary driver that allows these DCBH haloes to form and remain star-free is dynamical heating achieved through the rapid growth of these haloes. The rapid growth is strongly correlated with overdense environments with 76 DCBH candidate haloes forming in the \rarepeak simulation and only 3 DCBH candidate haloes forming in the \normal region. We also note that rapid growth by itself does not guarantee that a halo will become a DCBH candidate. Successfully avoiding metal enrichment must also be accounted for. Hence, in order to derive
an accurate sub-grid prescription it will be necessary to account for genetic\footnote{Genetic metal pollution was initially coined by \cite{Dijkstra2014a} and refers to the transfer of metals from smaller to larger haloes via mergers and accretion.} metal pollution \citep{Schneider_2006, Dijkstra2014a}. We therefore note that only hydrodynamic simulations which self-consistently follow metal transport will be able to successfully identify DCBH candidates in this case. Prescriptions which attempt to identify DCBH candidates only through the rapid growth of (dark matter) haloes will over-estimate the number density of DCBH candidates unless a metal enrichment/transport method is also used which can identify genetic metal enrichment. It should also be noted that sufficient particle (mass) resolution will also be paramount to resolve bursts of 
accretion which can delay \molH formation for at least a sound crossing time \citep{Wise_2019}.

While less than $ 5 \%$ of DCBH candidate haloes are exposed to LW intensities of greater than 2 $J_{21}$ these are nonetheless the candidate haloes which display complete isothermal collapse. In the vast majority of cases our examination of the radial profiles of these DCBH candidate haloes show that the central core of the haloes cools due to the \molHc. The haloes that collapse isothermally are stronger candidates for forming a super-massive star while those which collapse non-isothermally still display rapid mass inflow these are more likely to form a dense stellar cluster \citep{Freitag_2006, Freitag_2008, Lupi_2014, Katz_2015}. However, it should be noted that the resolution, and subgrid physics modules, of Renaissance are not sufficient to probe the further evolution of these haloes. The formation of a supermassive star, a normal population of metal-free free stars and/or a dense stellar cluster may be the final outcome. In order to fully understand the further evolution of these systems we are now running zoom-in
  simulations across a handful of interesting haloes in order to undercover the next stage of evolution of these haloes.

Finally, our analysis also reveals the existence of five synchronous haloes with separations of between 200 pc and 500 pc on the cusp of undergoing collapse. These haloes represent excellent candidates for further investigation of the synchronised pair scenario \citep{Dijkstra_2008, Visbal_2014, Regan_2017}. Imminent star formation in one of the haloes will result in the adjacent haloes being subject to intense LW radiation which will prevent the adjacent halo from cooling due to \molHc. In that case the adjacent halo will remain on the atomic cooling track and will be a strong candidate for super-massive star formation. In addition to this the subsequent merger of the two haloes should provide a plentiful supply of baryonic matter with which to successfully generate a massive black hole seed. Zoom-in simulations of a number of promising DCBH candidate haloes are now underway.

\section*{Acknowledgements}
JHW was supported by NSF awards AST-1614333 and OAC-1835213, NASA grant NNX17AG23G, and Hubble theory grant HST-AR-14326.
BWO was supported in part by NSF awards PHY-1430152, AST-1514700, OAC-1835213, by NASA grants NNX12AC98G, NNX15AP39G, and by Hubble theory Grants HST-AR-13261.01-A and HST-AR-14315.001-A.  MLN was supported by NSF grants AST-1109243, AST-1615858, and OAC-1835213. 
The Renaissance simulations were 
performed on Blue Waters operated by the National Center for Supercomputing Applications (NCSA) with PRAC allocation support by the NSF (awards ACI-0832662, ACI-1238993, ACI-1514580).  This research is part of the
Blue Waters sustained-petascale computing project, which is supported by the NSF (awards OCI-0725070, ACI-1238993) and the state of Illinois. Blue Waters is a joint effort of the University of Illinois at Urbana-Champaign and its NCSA. We thank an anonymous referee whose comments greatly improved the clarify of the manuscript. The freely available astrophysical analysis code \texttt{yt}\citep{YT} and plotting library matplotlib was used to construct numerous plots within this paper. Computations described in this work were performed using the publicly-available {\sc Enzo} code, which is the product of a collaborative effort of many independent scientists from numerous institutions around the world.

\bibliographystyle{mn2e_w}
\bibliography{./mybib}
\end{document}